\PassOptionsToPackage{unicode}{hyperref}
\PassOptionsToPackage{hyphens}{url}
\PassOptionsToPackage{dvipsnames,svgnames,x11names}{xcolor}
\documentclass[
  12pt]{article}

\usepackage{amsmath,amssymb}

\usepackage{amsthm}   
\usepackage{bbm}     

\theoremstyle{plain}
\newtheorem{theorem}{Theorem}
\newtheorem{lemma}[theorem]{Lemma}
\newtheorem{condition}{Condition}

\theoremstyle{definition}

\newtheorem{remark}[theorem]{Remark}

\usepackage[ruled, vlined, linesnumbered]{algorithm2e}
\SetAlgoNoLine
\SetKwFor{For}{for}{do}{end for}
\SetKwIF{If}{ElseIf}{Else}{if}{then}{else if}{else}{end if}
\SetKwInput{KwIn}{Input}
\SetKwInput{KwOut}{Output}
\SetKw{KwTo}{to}

\usepackage{iftex}
\ifPDFTeX
  \usepackage[T1]{fontenc}
  \usepackage[utf8]{inputenc}
  \usepackage{textcomp} 
\else 
  \usepackage{unicode-math}
  \defaultfontfeatures{Scale=MatchLowercase}
  \defaultfontfeatures[\rmfamily]{Ligatures=TeX,Scale=1}
\fi
\usepackage{lmodern}
\ifPDFTeX\else  
\fi
\IfFileExists{upquote.sty}{\usepackage{upquote}}{}
\IfFileExists{microtype.sty}{
  \usepackage[]{microtype}
  \UseMicrotypeSet[protrusion]{basicmath} 
}{}
\makeatletter
\@ifundefined{KOMAClassName}{
  \IfFileExists{parskip.sty}{%
    \usepackage{parskip}
  }{
    \setlength{\parindent}{0pt}
    \setlength{\parskip}{6pt plus 2pt minus 1pt}}
}{
  \KOMAoptions{parskip=half}}
\makeatother
\usepackage{xcolor}
\makeatletter
\ifx\paragraph\undefined\else
  \let\oldparagraph\paragraph
  \renewcommand{\paragraph}{
    \@ifstar
      \xxxParagraphStar
      \xxxParagraphNoStar
  }
  \newcommand{\xxxParagraphStar}[1]{\oldparagraph*{#1}\mbox{}}
  \newcommand{\xxxParagraphNoStar}[1]{\oldparagraph{#1}\mbox{}}
\fi
\ifx\subparagraph\undefined\else
  \let\oldsubparagraph\subparagraph
  \renewcommand{\subparagraph}{
    \@ifstar
      \xxxSubParagraphStar
      \xxxSubParagraphNoStar
  }
  \newcommand{\xxxSubParagraphStar}[1]{\oldsubparagraph*{#1}\mbox{}}
  \newcommand{\xxxSubParagraphNoStar}[1]{\oldsubparagraph{#1}\mbox{}}
\fi
\makeatother

\usepackage{longtable,booktabs,array}
\usepackage{calc} 
\usepackage{etoolbox}
\makeatletter
\patchcmd\longtable{\par}{\if@noskipsec\mbox{}\fi\par}{}{}
\makeatother
\IfFileExists{footnotehyper.sty}{\usepackage{footnotehyper}}{\usepackage{footnote}}
\makesavenoteenv{longtable}
\usepackage{graphicx}
\makeatletter
\def\maxwidth{\ifdim\Gin@nat@width>\linewidth\linewidth\else\Gin@nat@width\fi}
\def\maxheight{\ifdim\Gin@nat@height>\textheight\textheight\else\Gin@nat@height\fi}
\makeatother
\setkeys{Gin}{width=\maxwidth,height=\maxheight,keepaspectratio}
\makeatletter
\def\fps@figure{htbp}
\makeatother

\makeatletter
\@ifpackageloaded{caption}{}{\usepackage{caption}}
\AtBeginDocument{%
\ifdefined\contentsname
  \renewcommand*\contentsname{Table of contents}
\else
  \newcommand\contentsname{Table of contents}
\fi
\ifdefined\listfigurename
  \renewcommand*\listfigurename{List of Figures}
\else
  \newcommand\listfigurename{List of Figures}
\fi
\ifdefined\listtablename
  \renewcommand*\listtablename{List of Tables}
\else
  \newcommand\listtablename{List of Tables}
\fi
\ifdefined\figurename
  \renewcommand*\figurename{Figure}
\else
  \newcommand\figurename{Figure}
\fi
\ifdefined\tablename
  \renewcommand*\tablename{Table}
\else
  \newcommand\tablename{Table}
\fi
}
\@ifpackageloaded{float}{}{\usepackage{float}}
\floatstyle{ruled}
\@ifundefined{c@chapter}{\newfloat{codelisting}{h}{lop}}{\newfloat{codelisting}{h}{lop}[chapter]}
\floatname{codelisting}{Listing}

\makeatother
\makeatletter
\@ifpackageloaded{caption}{}{\usepackage{caption}}
\@ifpackageloaded{subcaption}{}{\usepackage{subcaption}}
\makeatother

\ifLuaTeX
  \usepackage{selnolig}  
\fi
\usepackage[]{natbib}
\setcitestyle{semicolon} 
\usepackage{bookmark}

\IfFileExists{xurl.sty}{\usepackage{xurl}}{} 
\hypersetup{
  pdftitle={Adaptive spatial blocking for scalable clustering inference with applications to high-throughput spatial proteomics},
  pdfauthor={Mingyu Go; Julia Wrobel; Hoseung Song},
  pdfkeywords={Ripley's K; Block averaging approaches; Spatial clustering; Permutation method; Large-scale images},
  colorlinks=true,
  linkcolor={blue},
  filecolor={Maroon},
  citecolor={Blue},
  urlcolor={Blue},
  pdfcreator={LaTeX via pandoc}}

\newcommand{\anon}{1}

\begin{document}

\def\spacingset#1{\renewcommand{\baselinestretch}%
{#1}\small\normalsize} \spacingset{1}

\if1\anon
{
  \title{\bf Adaptive spatial blocking for scalable clustering inference with applications to high-throughput spatial proteomics}
  
  \author{Mingyu Go \\
    Graduate School of Data Science, KAIST \\
    Julia Wrobel \\
    Department of Biostatistics and Bioinformatics, Emory University \\
    and \\
    Hoseung Song\thanks{The author acknowledges the support of the National Research Foundation of Korea (NRF) grant funded by the Korea government(MSIT) (RS-2022-NR068758 and RS-2026-25471551).}\hspace{.2cm}\\
    Department of Industrial and Systems Engineering, KAIST}
  
  \maketitle
} \fi

\if0\anon
{
  \bigskip
  \bigskip
  \bigskip
  \begin{center}
    {\LARGE\bf Adaptive spatial blocking for scalable clustering inference with applications to high-throughput spatial proteomics}
  \end{center}
  \medskip
} \fi

\bigskip
\begin{abstract}
Ripley’s $K$-function is a widely used spatial summary statistic for assessing clustering in point patterns. However, existing $K$-based methods can be computationally prohibitive for large-scale data, particularly in high-throughput spatial proteomics, because they rely on spatial information from all points in the image. To address this challenge, we propose a computationally efficient block-based testing framework that extracts disjoint local blocks from an image and aggregates clustering evidence across them. The proposed adaptive spatial blocking algorithm constructs blocks satisfying point-count and shape constraints, enabling scalable spatial clustering inference and fast $p$-value computation through an asymptotic normal approximation. Numerical studies demonstrate that the proposed method provides a favorable balance between statistical power and computational efficiency. In an application to healthy human intestine spatial proteomics data, our method detects strong spatial aggregation of plasma cells and colocalization between plasma cells and macrophages, while scaling favorably to large images. 
\end{abstract}

\noindent%
{\it Keywords:} Ripley's $K$-function; Block averaging approach; Spatial clustering; Permutation method; Large-scale images
\vfill

\newpage
\spacingset{1.8} 

\section{Introduction}

The analysis of point patterns has gained prominence in contemporary statistics, driven by its broad applications across diverse fields ranging from epidemiology and biology to spatial economics and environmental science. Detecting and characterizing spatial clustering in point locations provide crucial insights into the underlying spatial processes and phenomena within specific domains of study. For instance, spatial clustering among immune cells in the tumor microenvironment is significantly associated with patient survival outcomes \citep{keren2018structured, wilson2022tumor, shen2024spatial}.

Ripley's $K$-function \citep{ripley1988statistical}, a cornerstone metric in point process analysis, is widely employed to quantify the spatial structure of point patterns. The $K$-function at radius $r$, denoted by $K(r)$, represents the expected number of points within distance $r$ from a typical point in the pattern, normalized by the overall intensity of the process.

Traditional spatial clustering tests rely on the assumption of stationarity, which underlies both the theoretical foundation of the $K$-function and the null hypothesis of complete spatial randomness (CSR). However, this assumption is frequently violated in practice, as spatial intensity often exhibits substantial variation across the observation window. Such inhomogeneity can lead to inflated $K$-function estimates and consequently result in false detection of spatial aggregation, undermining the reliability of clustering tests.

To address the challenges posed by spatial inhomogeneity, an alternative approach leverages background points---defined as all remaining points in the pattern excluding the point type of primary interest. By employing label permutation across all points within the observation window, the empirical null distribution of the $K$-function can be estimated \citep{wilson2022tumor}. Under this label permutation framework, spatial clustering can be redefined as relative clustering with respect to the background point distribution. Given $n$ points with $m$ points of interest, the corresponding null hypothesis for clustering tests becomes that the intensity of points of interest, $\lambda_m(u)$, is proportional to that of the background points, $\lambda_{n-m}(u)$, such that $\lambda_m(u) \propto \lambda_{n-m}(u)$ at any location $u$ within the observation window.

Despite its theoretical appeal and practical advantages in handling inhomogeneous point patterns, the permutation-based approach faces substantial computational challenges when applied to large-scale datasets. The computational burden grows rapidly with both the number of points and the number of permutation replicates required for reliable statistical inference, making the method prohibitively expensive for contemporary large-scale spatial applications.

\subsection{The $KAMP$ framework and its limitations}
\label{sec:background}

\begin{figure}[h!]
    \centering
    \includegraphics[width=5.5in]{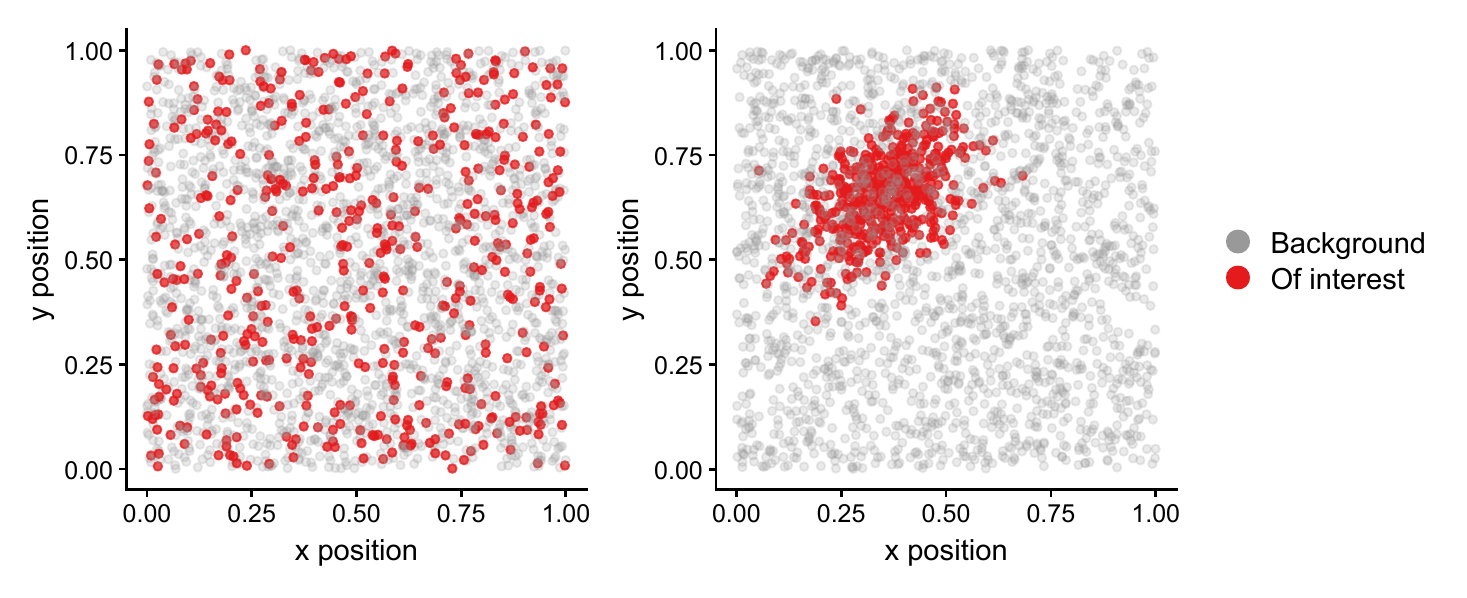}
    \caption{Simulated point patterns under the null (left) and alternative (right) hypotheses.}
    \label{fig:hypothesis}
\end{figure}

To address the computational burden of traditional permutation tests while maintaining the flexibility to handle inhomogeneous point patterns, \citet{wrobel2024robust} recently proposed the $KAMP$ framework ($K$ adjustment by Analytical Moments of the Permutation distribution). This approach tests the null hypothesis of proportional intensity against the alternative hypothesis of spatial clustering; a departure from proportional intensity indicating relative clustering. Figure~\ref{fig:hypothesis} illustrates simulated point patterns on the unit square $[0,1]^2$ under the null and alternative hypotheses. 

To test these hypotheses, for a spatial point pattern containing $n$ total points with $m$ points of interest, where $p = m/n$ denotes the proportion of points of interest, the $KAMP$ framework operates by permuting point labels while preserving their spatial locations. Let $\hat{K}(r)$ denote the empirical $K$-function estimator at radius $r$ for the points of interest. Under label permutation, $KAMP$ derives the closed-form analytical expressions for its expectation $E(\hat{K}(r))$ and variance $Var(\hat{K}(r))$ under the null hypothesis. Moreover, under the permutation null distribution, the standardized test statistic $Z(r)$ converges in distribution to a standard normal distribution as $n \to \infty$ and $m/n \to p \in (0,1)$:
\begin{equation}
    Z(r) = \frac{\hat{K}(r) - E(\hat{K}(r))}{\sqrt{Var(\hat{K}(r))}} \xrightarrow{d} \mathcal{N}(0,1).
    \label{eq:z_statistic}
\end{equation}

\noindent The analytical expressions for $E(\hat{K}(r))$ and $Var(\hat{K}(r))$ involve summations over all pairwise distances between points, requiring $O(n^2)$ distance calculations. For detailed regularity conditions and explicit formulas, we refer to \citet{wrobel2024robust}. This theoretical foundation eliminates the need for a large number of actual permutation replicates, enabling rapid $p$-value computation through direct evaluation of the normal distribution rather than empirical quantile estimation, as in classical permutation tests.

While the $KAMP$ framework offers a substantial speedup over standard permutation tests, it still faces significant computational limitations. We evaluate the execution times of $KAMP$ and the traditional permutation procedure in \texttt{R} on a 3.7 GHz AMD Ryzen 5 7500F 6-Core processor. Table~\ref{tab:runtimes} presents the average runtimes for computing the degree of clustering of points of interest, $\hat{K}(r)-E(\hat{K}(r))$, based on 10 simulated realizations of a homogeneous Poisson process with $n$ total points and $m=0.1n$ points of interest. For the permutation method, $E(\hat{K}(r))$ is estimated by averaging the $\hat{K}(r)$ values obtained from 1,000 permutations. 
\begin{table}[h!]
\centering
\caption{Average runtimes in seconds from 10 simulated point patterns with $n$ total points and $m=n/10$ points of interest.}
\label{tab:runtimes}
\begin{tabular}{@{}lrr@{}}
\toprule
$n$ & $KAMP$ & Permutation \\
\midrule            
10{,}000 &   8.19 &    67.50 \\
15{,}000 &  15.18 &   137.94 \\
20{,}000 &  24.54 &   230.04 \\
25{,}000 &  58.35 &   383.02 \\
30{,}000 &   -- &    819.65 \\
\bottomrule
\end{tabular}
\end{table}

The results indicate that, although $KAMP$ is substantially faster than the permutation method, it encounters severe memory constraints (indicated by ``--'' in Table~\ref{tab:runtimes}) because it requires storing pairwise distances among all points, and demands computational complexity $O(n^2)$. This issue is not specific to $KAMP$, but is inherent to Ripley's $K$-based analyses generally, since they rely on repeated pairwise distance calculations across spatial scales. This bottleneck is particularly critical in the analysis of high-throughput images, such as large-scale spatial proteomics data, which routinely consist of hundreds of thousands to millions of cells \citep{he2022high, jhaveri2023mapping}. 

To mitigate this burden, \citet{wrobel2024robust} proposed a subsampling strategy via independent thinning as a computationally feasible alternative. This approach reduces the sample size by performing independent Bernoulli trials with a retention probability $\tilde{p}$ for each point and then applies the $KAMP$ framework to the thinned sample. Although thinning improves computational efficiency, it results in significant loss of spatial information and may therefore lead to a pronounced decrease in statistical power. Moreover, independent thinning with a fixed $\tilde{p}$ yields only a constant-factor reduction in computation time and does not fundamentally resolve the underlying quadratic complexity.

\subsection{Our contributions}

To overcome the challenges discussed above, we propose block-based \textit{KAMP}, \textit{B-KAMP}, a spatial clustering test built upon block averaging approaches. The blocking strategy substantially reduces computational complexity, thereby improving scalability for high-throughput data \citep{zaremba2013b, song2021fast}. Furthermore, our block-based clustering test employs $KAMP$'s core statistic within each block to effectively detect local clustering signals.

The main contributions of this work are as follows: 

\begin{itemize}
 
    \item We propose an adaptive spatial blocking algorithm that identifies a set of rectangular blocks within a given observation window using an $R \times C$ grid representation. The algorithm efficiently determines a valid block configuration in $O(RC)$ time, enabling fast preprocessing even for large-scale images and local spatial analysis.
 
    \item We develop a block-based spatial clustering test for large-scale spatial point patterns. By replacing a full-window pairwise computation with block-level computations and aggregating the resulting local evidence through a weighted test statistic, the proposed test substantially improves scalability while retaining sensitivity to local clustering signals.
 
    \item We develop a general spatial blocking algorithm that enables efficient analysis of large-scale image data. While this paper focuses on spatial clustering tests, the proposed blocking strategy can be integrated more broadly into other scalable methods for massive spatial data. The codes for implementing the proposed method are available at GitHub (\url{https://github.com/mingyugo/B_KAMP}).

\end{itemize}

The remainder of this paper is organized as follows. Section~\ref{sec:methods} introduces the spatial blocking algorithm and the block-based framework for spatial clustering tests. Section~\ref{sec:simulation} evaluates the performance of the proposed method through extensive simulations, and Section~\ref{sec:real} demonstrates its practical utility using large-scale spatial proteomics data. We conclude with a discussion in Section~\ref{sec:discussion}.

\section{Methods}
\label{sec:methods}
In this section, we present a block-based framework for spatial clustering inference for large-scale spatial data. Section~\ref{sec:algorithm} introduces an adaptive spatial blocking algorithm that identifies a collection of blocks within a given observation window to support scalable and reliable spatial clustering testing. Section~\ref{sec:statistic} proposes a test statistic defined as a weighted sum of block-level clustering signals and a testing procedure motivated by a central limit theorem. Section~\ref{sec:bivariate} extends the framework to bivariate spatial clustering inference.

\subsection{Adaptive spatial blocking algorithm}
\label{sec:algorithm}

The spatial blocking algorithm is designed to achieve two goals: ensuring that each block is suitable for reliable spatial inference and identifying such blocks efficiently. To accomplish this, we first define block constraints that specify the geometric and statistical conditions required for a valid block. We then introduce a block identification algorithm that, among all candidate blocks satisfying these constraints, selects the one with the smallest total number of points. Finally, we present an adaptive spatial blocking algorithm that combines these components into a complete procedure for extracting a collection of disjoint blocks from the observation window.

\subsubsection{Block constraints}
\label{sec:constraints}

Reliable estimation of spatial summary statistics, such as Ripley's $K$, depends strongly on the geometry of the observation window. In spatial point pattern analysis, departures from standard rectangular shapes can introduce significant bias into edge-corrected estimators. In particular, \citet{goreaud1999explicit} emphasized that complex study windows (e.g., irregular polygons) complicate edge-effect correction and may induce artificial spatial heterogeneity, potentially leading to inflated $K$-function estimates or spurious clustering signals. Accordingly, rectangular windows are commonly adopted in spatial point pattern analysis, since many edge-correction methods are designed for such regular geometries \citep{haase1995spatial, pelissier2001practical}. Motivated by these considerations, we require each block to be a rectangular subregion of the observation window.

This rectangular shape alone, however, is not sufficient to ensure statistical stability. Rectangular blocks with large aspect ratios, i.e., highly elongated blocks, can pose challenges similar to those arising from irregular geometries. In particular, \citet{doguwa1989edge} showed that $K$-based estimators computed on highly elongated rectangular windows exhibit significantly larger mean squared errors than those computed on square windows, increasing the risk of misrepresenting the underlying spatial process. Consequently, to promote reliable $K$-function estimation, our algorithm imposes an upper bound on the aspect ratio, defined as the ratio of the longer side to the shorter side, so that selected blocks remain as square-like as possible.

In addition to these geometric considerations, the number of points within each block plays an important role in balancing statistical power and computational efficiency in the block-based testing framework developed in  Section~\ref{sec:statistic}. Blocks with too few observations may fail to capture local spatial structure, resulting in reduced power for detecting clustering. On the other hand, blocks containing too many points lead to computational complexity close to $O(n^2)$, limiting scalability for large datasets. In related block-averaging methods for large-scale testing, \citet{zaremba2013b} proposed choosing block sizes of order $\sqrt{n}$ as a practical compromise between statistical power and computational efficiency, reducing pairwise computational cost from $O(n^2)$ to $O(n^{1.5})$.

Motivated by these geometric and statistical considerations, we define a block as a rectangular subregion of the observation window that satisfies the following conditions:

\begin{enumerate}
    \item \textbf{Minimum block size constraint:}
    The block must contain at least $\sqrt{m}$ points of interest, where $m$ denotes the total number of points of interest in the observation window.
    
    \item \textbf{Maximum aspect ratio constraint:}
    The aspect ratio of the block must not exceed a prespecified threshold $\rho$.
\end{enumerate}

For the label-permutation approach used in our tests, we impose an additional lower bound on the number of background points, that is, the remaining $n-m$ points. Specifically, each block must contain at least $\min(\sqrt{m}, \sqrt{n-m})$ background points. This condition helps ensure a non-degenerate permutation distribution of $\hat{K}(r)$ within each block. Without sufficient background points, most permutations yield similar label configurations, causing the permutation distribution to concentrate near the observed value and thereby reducing statistical power. We discuss the choice of the aspect ratio $\rho$ in Section~\ref{sec:total_procedure}.

\subsubsection{Block identification algorithm}
\label{sec:block_detection}
 
A variety of geometric algorithms have been developed to identify rectangular or square subregions that satisfy point-count constraints, such as the minimum block size requirement described in Section~\ref{sec:constraints}. For example, the problem of finding a rectangle, square, or disc containing at least $k$ points has been studied extensively in computational geometry \citep{khanteimouri2013computing, barba2014k}. Beyond these enclosing problems, spatial indexing and adaptive partitioning schemes, such as quadtrees and their variants, provide efficient hierarchical decompositions of large point patterns into rectangular regions for accelerated spatial querying \citep{finkel1974quad, samet1984quadtree}.
 
These existing methods, however, are not directly suited to our block-identification setting. Geometric enclosing algorithms typically have superlinear computational complexity, ranging from $O(n \log n)$ to $O(n^2)$, which becomes burdensome for large-scale spatial point patterns. In the case of quadtree-based partitioning schemes, the observation window is recursively subdivided according to geometric criteria, without explicitly enforcing the minimum point-count constraints required in our setting. Consequently, these approaches are not directly applicable to our adaptive block identification framework.

To address these limitations, we adopt a grid-based approach inspired by the \textit{largest-rectangle-in-a-histogram} algorithm. Unlike the geometric methods discussed above, which depend superlinearly on the number of points $n$, this framework operates with computational complexity $O(RC)$ for a given $R \times C$ grid, making it highly efficient for identifying rectangular subregions in large-scale spatial patterns. The procedure begins by discretizing the observation window into an $R \times C$ grid and representing it as a binary matrix $G \in \{0,1\}^{R \times C}$, initialized with all ones, where $G_{ij} = 1$ indicates that the corresponding grid cell is available for block construction. 

\begin{figure}[h!]
    \centering
    \includegraphics[width=1\textwidth]{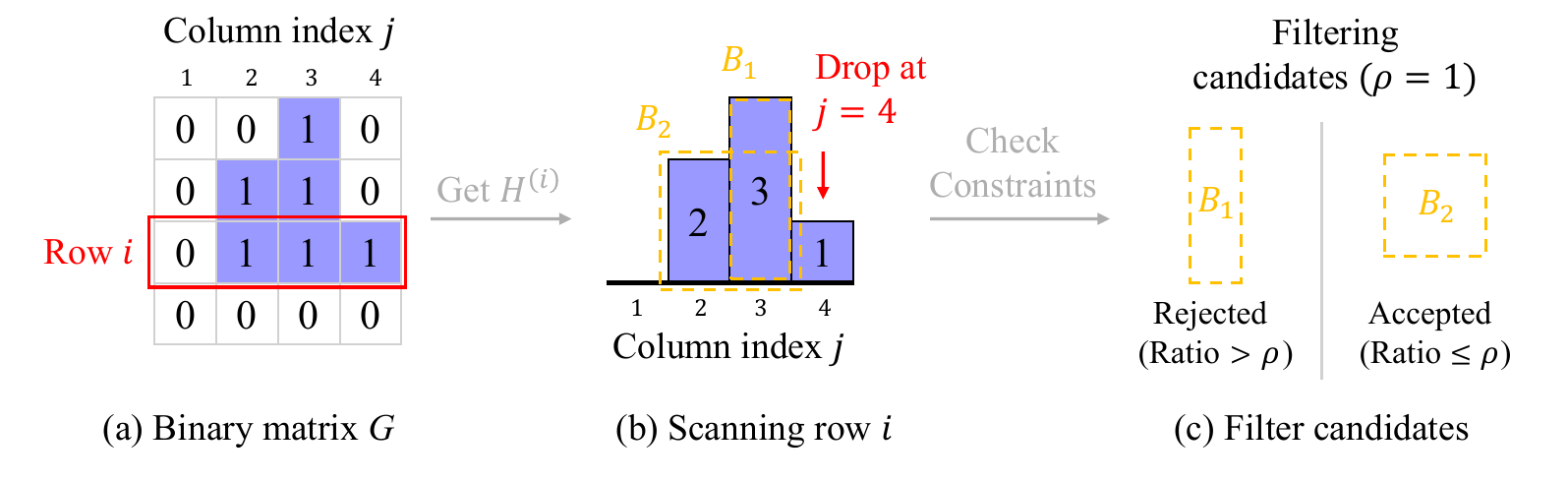}
    \caption{Illustration of extracting block candidates. (a) A binary matrix $G$ with row $i$ highlighted. (b) Scanning the histogram: a height drop at $j=4$ (from $3$ to $1$) triggers the exploration of maximal rectangles formed by preceding columns (orange dashed box). (c) Evaluating candidates: under the aspect ratio constraint $\rho = 1$, the $3 \times 1$ rectangle ($B_1$) is rejected, while the $2 \times 2$ shape ($B_2$) is retained as a block candidate.}
    \label{fig:algorithm_vis}
\end{figure}

To efficiently identify block candidates, we use a row-wise histogram representation of $G$. Specifically, for each row $i$, we construct a height vector $H^{(i)}$ whose entries record the number of consecutive available cells ending at row $i$. Algorithm~\ref{alg:build_histogram} computes these height vectors and serves as a preprocessing step for Algorithm~\ref{alg:block_selection}, which uses them to extract block candidates. The overall block identification procedure is illustrated in Figure~\ref{fig:algorithm_vis} and summarized as follows:
\begin{algorithm}[h!]
\small
\caption{Build Histogram at Each Row for a Binary Matrix $G$}
\label{alg:build_histogram}

\KwIn{A binary matrix $G \in \{0,1\}^{R \times C}$.}
\KwOut{Height vectors $\{H^{(i)}\}_{i=1}^{R}$, where $H^{(i)} \in \mathbb{Z}_{\ge 0}^{C}$.}

Initialize $H^{(0)}[j] \leftarrow 0$ for all $j \in \{1, \dots, C\}$\; 

\For{$i \leftarrow 1$ \KwTo $R$}{
    \For{$j \leftarrow 1$ \KwTo $C$}{
        \If{$G[i,j] = 1$}{
            $H^{(i)}[j] \leftarrow H^{(i-1)}[j] + 1$\;
        }
        \Else{
            $H^{(i)}[j] \leftarrow 0$\;
        }
    }
}
\end{algorithm}

\begin{algorithm}[h!]
\small
\SetArgSty{textnormal} 
\SetKwComment{Comment}{}{}
\SetKw{KwTo}{to} 

\caption{Block Identification from $G$}
\label{alg:block_selection}

\KwIn{A binary matrix $G \in \{0,1\}^{R \times C}$.}
\KwOut{A selected block \texttt{block} and an updated matrix $G$.}

Initialize $S \leftarrow \emptyset$ and $T \leftarrow \emptyset$\;

Compute height vectors $\{H^{(i)}\}_{i=1}^{R}$ via Algorithm~\ref{alg:build_histogram}\;

\For{$i \leftarrow 1$ \KwTo $R$}{
    $T \leftarrow \emptyset$\;

    \For{$j \leftarrow 1$ \KwTo $C+1$}{
        \If{$j \le C$}{
            $h \leftarrow H^{(i)}[j]$\;
        }
        \Else{
            $h \leftarrow 0$\;
        }

        $idx \leftarrow j$\;

        \tcp{Exploration of maximal rectangles}
        \While{there exists $t = (t_1, t_2) \in T$ such that $t_2 \ge h$}{
            Select $t \in T$ with the maximum $t_2$\;
            $B \leftarrow G[i - t_2 + 1 : i,\; t_1 : j-1]$\tcp*[r]{$B$: maximal rectangle}
            \If{$B$ satisfies block constraints}{
                $S \leftarrow S \cup \{B\}$\tcp*[r]{$B$: block candidate}
            }
            $idx \leftarrow t_1$\; 
            $T \leftarrow T \setminus \{t\}$\;
        }
        $T \leftarrow T \cup \{(idx,\, h)\}$\;
    }
}

Select $\texttt{block} = G[r_1:r_2,\; c_1:c_2]$ from $S$ with the minimum total number of points\;

Update $G$: $G_{u,v} \leftarrow 0$ for all $u \in \{r_1, \dots, r_2\}$, $v \in \{c_1, \dots, c_2\}$\;

\Return{$(\texttt{block},\; G)$}\;
\end{algorithm}

\begin{enumerate}
    \item Construct the histogram $H^{(i)}$ for row $i$. For example, in Figure~\ref{fig:algorithm_vis}(a), when $i=3$, the corresponding heights are $0,2,3,1$.
    
    \item Scan the columns of row $i$. A drop in height, such as the transition from $3$ to $1$ at $j=4$ in Figure~\ref{fig:algorithm_vis}(b), triggers the maximal-rectangle exploration step in Algorithm~\ref{alg:block_selection}.

    \item For each identified maximal rectangle, check whether it satisfies the minimum block size and maximum aspect ratio ($\rho$) constraints. For example, in Figure~\ref{fig:algorithm_vis}(c), with $\rho=1$, $B_2$ is accepted as a block candidate while $B_1$ is rejected for exceeding the ratio.
        
    \item Repeat Steps 1--3 for all rows $i$ to collect the full set of block candidates $S$.
    
    \item Select the final \texttt{block} from $S$ that contains the smallest total number of points.
    
    \item Update $G$ by setting the grid cells covered by the selected \texttt{block} to $0$.
\end{enumerate}
 
The selection criterion in Step~5 is particularly important. By prioritizing the candidate block with the smallest total number of points, the algorithm leaves more available space in $G$ for subsequent block identification, thereby increasing the total number of disjoint blocks that can be identified. This selection rule tends to avoid excessively large blocks and to keep blockwise point counts relatively small and balanced, improving computational scalability and supporting the asymptotic normal approximation used in Section~\ref{sec:statistic}.

\subsubsection{Adaptive spatial blocking algorithm}
\label{sec:total_procedure}

\begin{figure}[h!]
    \centering
    \begin{subfigure}[b]{0.45\textwidth}
        \centering
        \includegraphics[width=\linewidth]{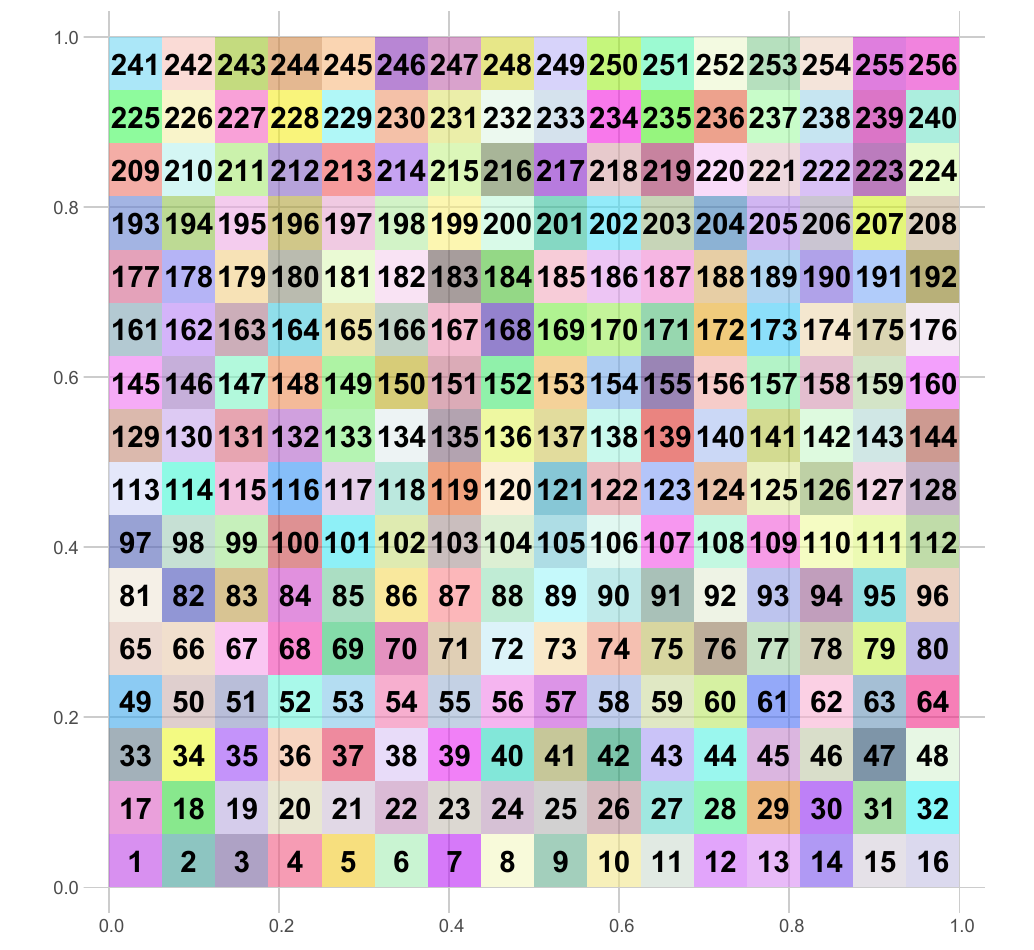}
        \caption{Grid representation}
        \label{fig:grid}
    \end{subfigure}
    \hfill
    \begin{subfigure}[b]{0.45\textwidth}
        \centering
        \includegraphics[width=\linewidth]{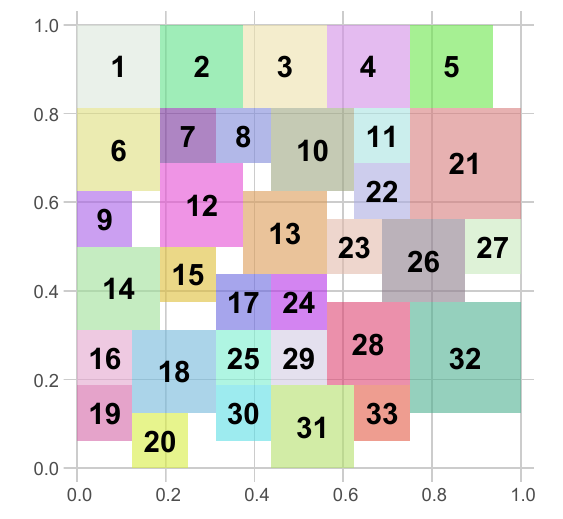}
        \caption{Phase~I}
        \label{fig:phase1}
    \end{subfigure}
    
    \vspace{0.2cm} 
    
    \begin{subfigure}[b]{0.45\textwidth}
        \centering
        \includegraphics[width=\linewidth]{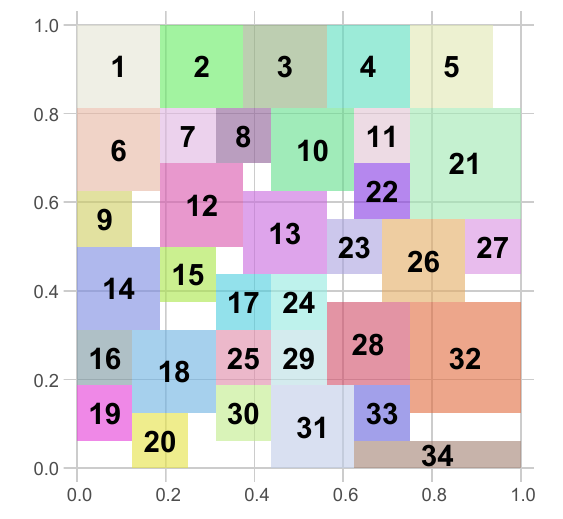}
        \caption{Phase~II}
        \label{fig:phase2}
    \end{subfigure}
    \hfill
    \begin{subfigure}[b]{0.45\textwidth}
        \centering
        \includegraphics[width=\linewidth]{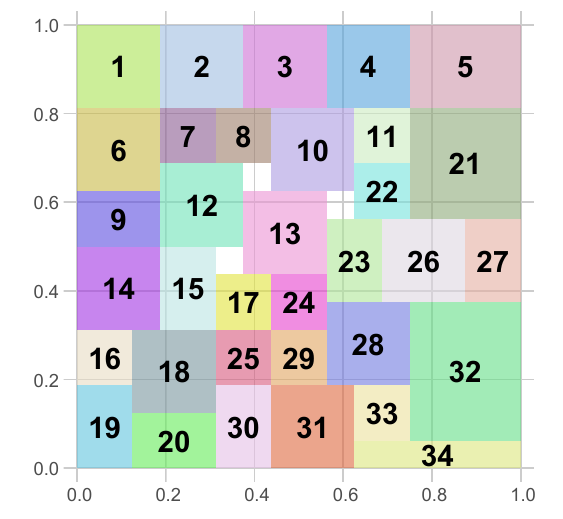}
        \caption{Phase III}
        \label{fig:complement}
    \end{subfigure}

    \caption{Result of applying the adaptive spatial blocking algorithm to a simulated point pattern with $\lambda_m = 4{,}000$ and $\lambda_{n-m} = 36{,}000$ on an observation window $[0,1]^2$. (a)~Grid representation of the point pattern; (b)~Phase~I with $\rho_1$ equal to the aspect ratio of the observation window (i.e., $\rho_1 = 1$); (c)~Phase~II with $\rho_2 = \infty$; (d)~Phase III, block complement.}
    \label{fig:process}
\end{figure}

Building on the block constraints in Section~\ref{sec:constraints} and block identification algorithm in Section~\ref{sec:block_detection}, we now describe the complete procedure for extracting a collection of disjoint rectangular blocks from a spatial point pattern. The proposed procedure consists of three sequential stages: (i) grid representation, (ii) block identification (Phases~I and~II), and (iii) block complement (Phase~III). An illustration of the entire procedure is provided in Figure~\ref{fig:process} using a simulated point pattern on the unit square $[0,1]^2$, generated by superimposing two independent homogeneous Poisson processes: one for the target points with intensity $\lambda_m = 4,000$ and the other for the background points with intensity $\lambda_{n-m} = 36,000$.

\begin{figure}[h!]
\centering
\includegraphics[width=0.85\textwidth]{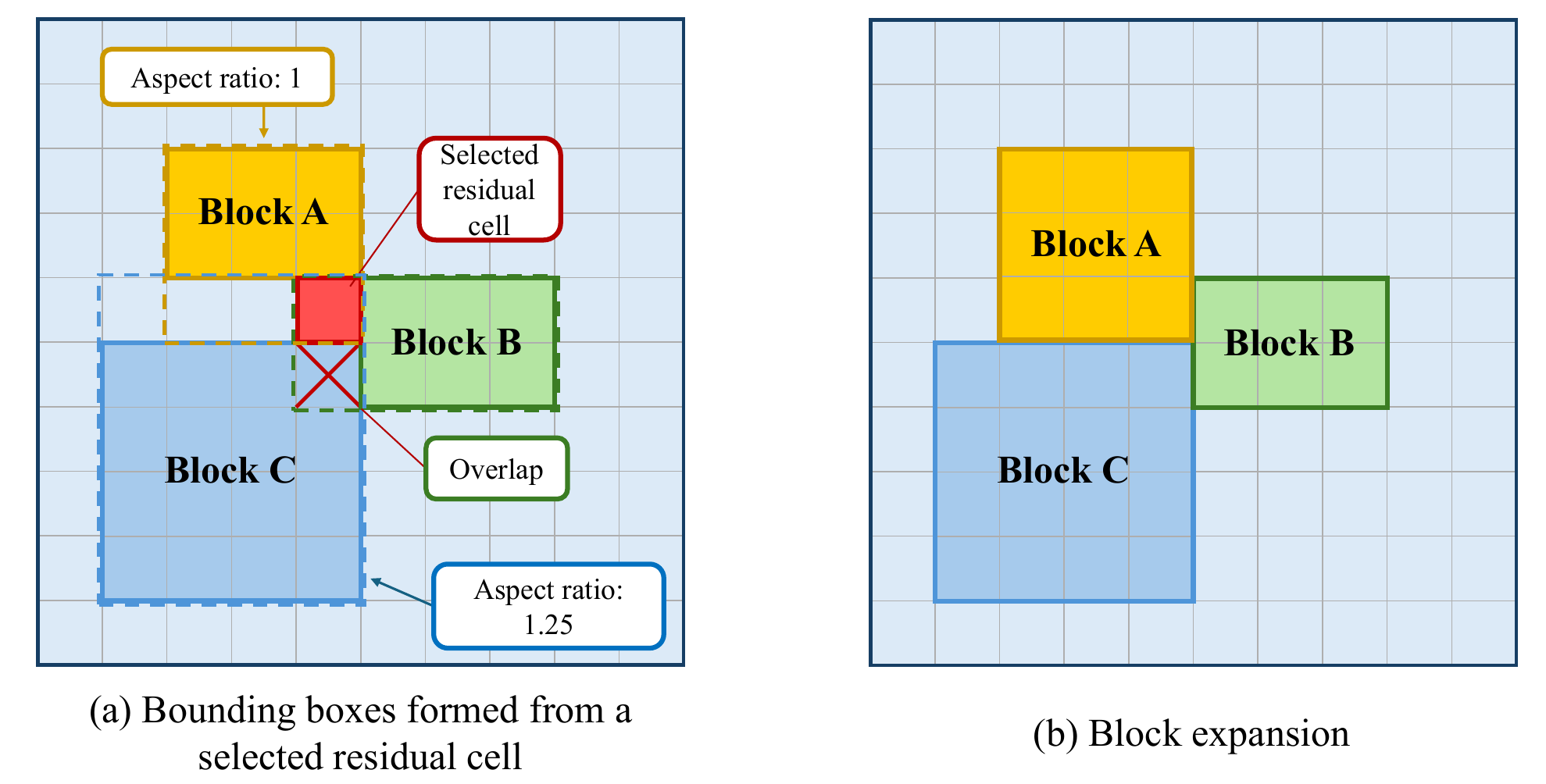}
\caption{Illustration of block complement (Phase III).}
\label{fig:block_complement}
\end{figure}

\begin{enumerate}
\item \textbf{Grid representation.}
We first construct an adaptive regular grid over the spatial domain of the point pattern by progressively increasing the grid resolution until each cell contains at most $\sqrt{n}$ points, resulting in equally sized rectangular cells (Figure~\ref{fig:process}(a)). This grid is then represented as a binary matrix $G \in \{0,1\}^{R \times C}$, where $G_{ij} = 1$ indicates that cell $(i,j)$ is available for block construction and $G_{ij} = 0$ indicates that it has already been assigned to a selected block. Initially, all entries of $G$ are set to 1. If an individual cell already satisfies the block constraints in Section~\ref{sec:constraints}, it is immediately treated as a block and the corresponding entry in $G$ is set to 0.

\item \textbf{Block identification (Phases I and II).}
Blocks are identified by repeatedly applying Algorithm~\ref{alg:block_selection} to $G$ under a prescribed aspect ratio constraint $\rho$. This stage consists of two phases, using aspect ratio thresholds $\rho_1$ and $\rho_2$ sequentially, where $\rho_1 < \rho_2$. Phase~I prioritizes the extraction of near-square blocks by imposing a strict aspect ratio constraint $\rho_1$ (e.g., in Figure~\ref{fig:process}(b), $\rho_1$ equals the aspect ratio of the observation window, which is 1). Algorithm~\ref{alg:block_selection} is applied iteratively until no further blocks satisfying this constraint can be identified.
In Phase~II, the aspect ratio constraint is relaxed to $\rho_2$ (e.g., in Figure~\ref{fig:process}(c), $\rho_2 = \infty$), allowing the identification of more elongated rectangular blocks. This increased flexibility facilitates the identification of additional blocks, thereby enhancing computational efficiency and strengthening the asymptotic result of the test statistic. Algorithm~\ref{alg:block_selection} is applied iteratively until no further blocks satisfying the relaxed constraint can be identified. Upon completion of both phases, some grid cells may remain unassigned (i.e., $G_{ij} = 1$); these residual cells are handled in Phase~III.

\item \textbf{Block Complement (Phase III).}
Phase III aims to incorporate residual cells remaining after Phase~II into existing blocks in order to reduce the loss of spatial coverage while preserving the disjointness of the block collection, as illustrated in Figure~\ref{fig:process}(d). Figure~\ref{fig:block_complement} provides a detailed illustration of this step. For each residual cell, we recognize all identified blocks that share a common edge with it, referred to as adjacent blocks. For each adjacent block, we construct the minimal bounding rectangle enclosing both the block and the residual cell. Among all such bounding rectangles that remain disjoint from all other blocks, we select the one with the smallest aspect ratio, breaking ties by choosing the rectangle containing the fewest total points. The corresponding block is then expanded to the selected bounding rectangle, and the grid matrix $G$ is updated accordingly. This procedure is repeated iteratively until no further expansions are possible.
\end{enumerate}

In Phase~I, we recommend setting $\rho_1$ equal to the aspect ratio of the observation window, so that it naturally accommodates general rectangular windows and extracts blocks that are as compact as possible in accordance with the window geometry. When $\rho_1$ is set in this way, Phase~I typically identifies most of the blocks, leaving only limited residual subregions for Phase~II. Furthermore, since Algorithm~\ref{alg:block_selection} selects, among all valid candidates, the block containing the fewest total points, the choice of $\rho_2$ has little practical impact on statistical performance and runtime once $\rho_1$ is appropriately specified. We therefore recommend setting $\rho_2 = \infty$ in Phase~II. Simulation results in Supplement~A also demonstrate that, with $\rho_1$ set to the aspect ratio of the observation window, empirical power and runtime vary negligibly across different values of $\rho_2$. Detailed sensitivity results are provided in Supplement~A.

The proposed blocking algorithm provides a scalable representation of spatial structures and can be readily incorporated into existing $K$-function-based analyses. This combination enables efficient analysis of large-scale image data, where the direct application of traditional methods would be computationally prohibitive.

\subsection{Test statistics}
\label{sec:statistic}
Applying the proposed adaptive spatial blocking algorithm in Section~\ref{sec:total_procedure} to a spatial point pattern yields $b$ disjoint blocks. For each resulting block $i$, let $n_i,\; m_i,$ and $|A_i|$ denote the total number of points, the number of points of interest, and the area of the block, respectively. Let $\hat{K}_i(r)$ be the estimator of Ripley's $K$ for the points of interest at radius $r$ in the $i$-th block. Specifically, $\hat{K}_i(r)$ is the normalized mean count of pairwise distances between points of interest that fall within radius $r$ in the $i$-th block. For $i=1, \dots, b$, $\hat{K}_i(r)$ is defined as 
\begin{equation}
\label{eq:ripley_k_block}
\hat{K}_i(r) = \frac{|A_i|}{m_i(m_i-1)} \sum_{u=1}^{m_i} \sum_{v \neq u}^{m_i} \mathbbm{1}(d\{c_u^{(i)},c_v^{(i)}\} \leq r) e_{uv}^{(i)},
\end{equation}
where $d\{c_u^{(i)},c_v^{(i)}\}$ is the distance between points $c_u^{(i)}$ and $c_v^{(i)}$ in the $i$-th block, $\mathbbm{1}(\cdot)$ is an indicator function, and $e_{uv}^{(i)}$ is the edge correction term to ensure unbiasedness within the $i$-th block. In our numerical studies, we employ translation edge correction \citep{baddeley2016spatial} to account for edge effects.

We build upon the $KAMP$ approach \citep{wrobel2024robust}, which accounts for spatial inhomogeneity through point label-permutation and enables fast $p$-value computation by deriving analytical expressions for the moments and asymptotic distribution under the permutation null distribution. Here, for the $i$-th block, we adopt the permutation null distribution, which assigns an equal probability of $1 / \binom{n_i}{m_i}$ to each of the $\binom{n_i}{m_i}$ possible assignments of $m_i$ points of interest among the $n_i$ total points. 

The standardized statistic for \eqref{eq:ripley_k_block}, which serves as the core statistic in our framework, is defined as
\begin{equation}
\label{eq:standardized_ripley_k}
Z_i(r) = \frac{\hat{K}_i(r) - E(\hat{K}_i(r))}{\sqrt{Var(\hat{K}_i(r))}}, 
\end{equation}

where $E(\hat{K}_i(r))$ and $Var(\hat{K}_i(r))$ are the expectation and variance of $\hat{K}_i(r)$ under the permutation null distribution, respectively. The statistic $Z_i(r)$ measures the clustering signal at radius $r$ within the $i$-th block, and a large value of $Z_i(r)$ provides evidence of significant spatial clustering. The analytical expressions for $E(\hat{K}_i(r))$ and $Var(\hat{K}_i(r))$, derived in a similar manner to \citet{wrobel2024robust}, are provided in Lemma \ref{lem:moments}. 

\begin{lemma} \label{lem:moments} Let $W_{uv}^{(i)}(r) = \mathbbm{1}(d\{c_u^{(i)},c_v^{(i)}\} \leq r) e_{uv}^{(i)}$ for $i=1, \dots, b.$ Under the permutation null distribution, the expectation and variance of $\hat{K}_i(r)$ for $i = 1, \dots, b$ are given by
\begin{align*}
E(\hat{K}_i(r))
&= \frac{|A_i|}{n_i(n_i-1)}\, R_0^{(i)}, \\
Var(\hat{K}_i(r))
&= \frac{|A_i|^2}{m_i^2(m_i-1)^2}
\left\{
2R_1^{(i)} p_{1,i}
+ 4R_2^{(i)} p_{2,i}
+ R_3^{(i)} p_{3,i}
\right\}
- E^2(\hat{K}_i(r)),
\end{align*}
where
\begin{align*}
p_{1,i} &= \frac{m_i(m_i-1)}{n_i(n_i-1)}, \quad
p_{2,i} = \frac{m_i(m_i-1)(m_i-2)}{n_i(n_i-1)(n_i-2)}, \quad
p_{3,i} = \frac{m_i(m_i-1)(m_i-2)(m_i-3)}{n_i(n_i-1)(n_i-2)(n_i-3)}, \\
R_0^{(i)}
&= \sum_{u=1}^{n_i} \sum_{\substack{v \neq u}}^{n_i}
W_{uv}^{(i)}(r), \quad
R_1^{(i)}
= \sum_{u=1}^{n_i} \sum_{\substack{v \neq u}}^{n_i}
\bigl(W_{uv}^{(i)}(r)\bigr)^2, \quad 
R_2^{(i)}
= \sum_{u=1}^{n_i} \sum_{\substack{v \neq u}}^{n_i}
\sum_{\substack{w \neq u \neq v}}^{n_i}
W_{uv}^{(i)}(r)\, W_{uw}^{(i)}(r), \\
R_3^{(i)}
&= \sum_{u=1}^{n_i} \sum_{\substack{v \neq u}}^{n_i}
\sum_{\substack{w \neq u \neq v}}^{n_i}
\sum_{\substack{ t \neq u \neq v \neq w}}^{n_i}
W_{uv}^{(i)}(r)\, W_{wt}^{(i)}(r).
\end{align*}

\end{lemma}

Based on these blockwise standardized statistics, we propose a new test statistic,
\begin{equation}
    Z(r) = \sum_{i=1}^{b} w_i Z_i(r),
\end{equation}
defined as a weighted sum of the blockwise standardized statistics. The weights are normalized so that the sum of their squared values equals one. Because each blockwise statistic $Z_{i}(r)$ is centered and variance-standardized under the conditional blockwise permutation null defined below, the choice of weights primarily affects power rather than null calibration. In particular, when the weights are functions of the realized block structure and blockwise counts, they are treated as fixed under the conditional calibration step. More generally, the asymptotic argument below applies to any such normalized weighting scheme, provided that no single block receives dominant weight.

As our default weighting scheme, we use the pre-specified choice in which $w_i\propto n_i/p_i$. This weighting scheme places greater emphasis on blocks containing more points and a lower local prevalence of the points of interest. The motivation is that, under sparse-signal alternatives, weak relative clustering may be harder to detect in low-abundance blocks and may therefore benefit from upweighting. A sensitivity analysis in Supplement~B compares several weighting schemes across different abundance levels and number of cluster centers. The result shows that weights incorporating $1/p_i$ generally improve power, with $w_i \propto n_i/p_i$ performing best among the schemes considered.

We next derive the asymptotic null distribution of $Z(r)$, which enables fast $p$-value computation. We condition on the block partition produced by the adaptive blocking algorithm, the point locations within each block, and the blockwise counts $\left\{(n_{i}, m_{i})\right\}_{i=1}^{b}$. Conditional on these quantities, the blockwise permutation null assigns equal probability to each of the $\binom{n_i}{m_i}$ relabelings of the $n_{i}$ points in block $i$ that preserve exactly $m_{i}$ points of interest. Relabelings are generated independently across blocks. Thus, under this conditional blockwise permutation null, $\{Z_i(r)\}_{i=1}^b$ are independent but not identically distributed.

\begin{condition} 
As $n\rightarrow\infty$ and $m/n\rightarrow p\in (0,1)$ with $b\rightarrow\infty$, $\max_{1\le i\le b}|w_{i}| \rightarrow 0$.
\label{cond:max_weight}
\end{condition}

\begin{condition} 
For a fixed radius $r$, $\sup_{1\leq i \leq b} E^{*}\{|Z_i(r)|^3\} < \infty$, where $E^{*}$ denotes expectation under the conditional blockwise permutation null.
\label{cond:boundedness} 
\end{condition}

Under Conditions \ref{cond:max_weight} and \ref{cond:boundedness}, as $n\rightarrow\infty$ and $m/n\rightarrow p\in (0,1)$, 
\begin{align*}
    Z(r) \stackrel{d}{\rightarrow} N(0,1)
\end{align*}
under the conditional blockwise permutation null.

This result follows from the Lyapunov CLT. Let $X_{i} = w_{i}Z_{i}(r)$. Under the conditional blockwise permutation null, $\{X_{i}\}_{i=1}^{b}$ are independent, $E^{*}(X_{i}) = 0$, and 
\begin{align*}
 \sum_{i=1}^{b}Var^{*}(X_{i}) = \sum_{i=1}^{b}w_{i}^2 = 1.
\end{align*}

Moreover,
\begin{align*}
    \sum_{i=1}^{b}E^{*}\{|X_{i}|^3\} \le \left(\max_{1\le i\le b}|w_{i}|\right)\left(\sup_{1\leq i \leq b} E^{*}\{|Z_i(r)|^3\}\right)\sum_{i=1}^{b}w_{i}^2 \rightarrow 0.
\end{align*}
Hence, the Lyapunov condition is satisfied.

\begin{remark}
    The proposed procedure is interpreted as a scalable conditional blockwise test. It conditions on the block partition and corresponds to an independent within-block relabeling scheme, rather than the full-window permutation distribution of the original $KAMP$ statistic.
\end{remark}

Condition \ref{cond:max_weight} is expected to hold when the adaptive blocking algorithm produces a reasonably balanced collection of blocks. For the default weights $w_i \propto n_i/p_i$, this condition is satisfied when the blockwise point counts are comparable and the local prevalences $p_i = m_i/n_i$ remain bounded away from 0 and 1, in which case $\max_{1\le i\le b}|w_{i}|$ is of order $b^{-1/2}$. Condition \ref{cond:boundedness} is a standard Lyapunov-type moment condition that rules out nearly degenerate blockwise statistics, which may arise from very small or highly imbalanced blocks.

\begin{figure}[h!] 
\centering
\includegraphics[width=0.8\linewidth]{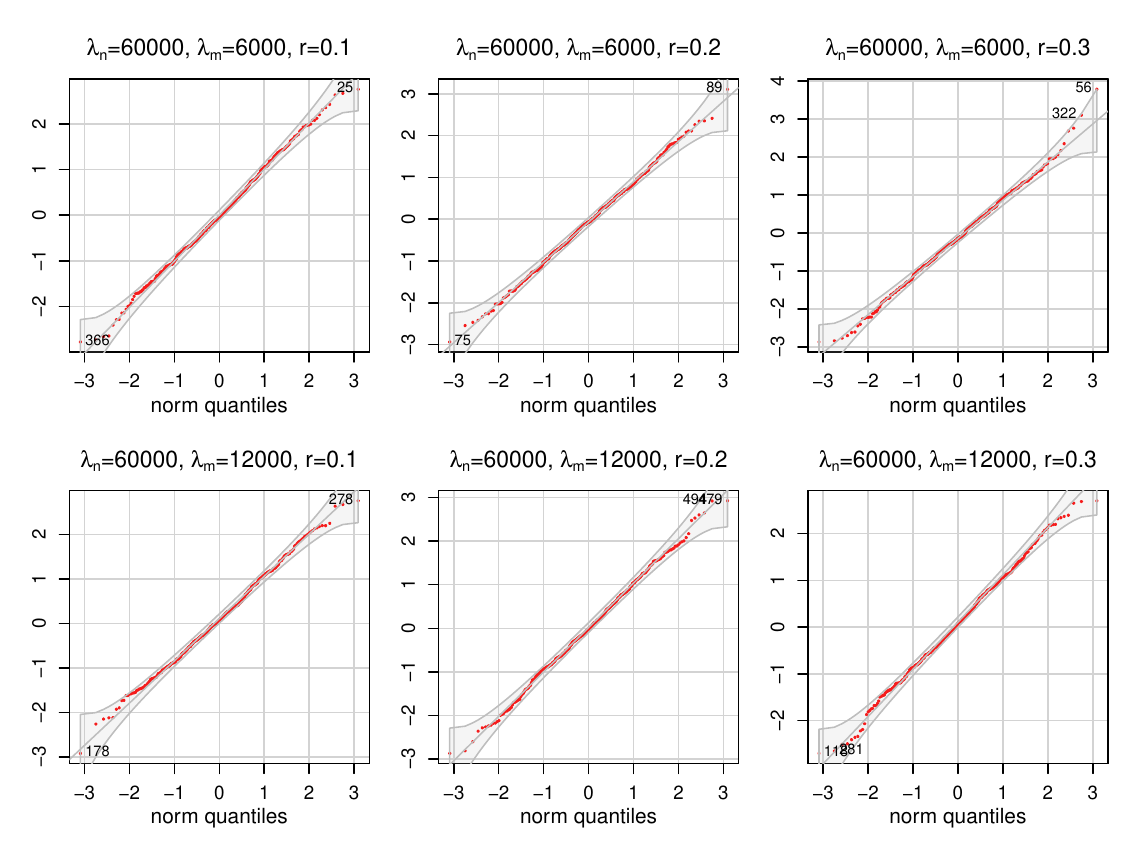}
\caption{Normal quantile-quantile plots of $Z(r)$.}
\label{fig:qqplot} 
\end{figure}

Figure \ref{fig:qqplot} displays the quantile-quantile plots of $Z(r)$ against the standard normal distribution under various $\lambda_m$ and $r$ with $\lambda_n$ fixed at $60{,}000$. Each value of $Z(r)$ is computed from $500$ realizations, each generated by superimposing two independent homogeneous Poisson point processes with intensities $\lambda_m$ and $\lambda_n - \lambda_m$ for points of interest and background points, respectively. The radius $r$ is taken to be proportional to the shorter side of each block. These plots suggest that the conditional Gaussian calibration is adequate in the simulated null settings considered here.

\subsection{Extension to bivariate spatial clustering}
\label{sec:bivariate}
While Section~\ref{sec:statistic} focuses on the spatial clustering of a single point type, spatial interactions between two point types are also of substantial interest in many practical applications. In spatial proteomics, for instance, quantifying the spatial colocalization of distinct immune cell types is a common objective \citep{schurch2020coordinated, samorodnitsky2024spatial, lee2025profiling}. The proposed block-based framework can be naturally extended to this bivariate setting.

To quantify spatial colocalization, we use a bivariate version of Ripley's $K$-function. Suppose that, within block $i$, there are $m_{i1}$ Type 1 points, $m_{i2}$ Type 2 points, and $n_i - (m_{i1} + m_{i2})$ background points. Analogously to the univariate estimator in equation \eqref{eq:ripley_k_block}, we define the block-specific bivariate $K$-function estimator as
\begin{equation}
\label{eq:biv_ripley_k_block}
\hat{K}_{12,i}(r) = \frac{|A_i|}{m_{i1}m_{i2}} \sum_{u=1}^{m_{i1}} \sum_{v=1}^{m_{i2}} \mathbbm{1}(d\{c_{1,u}^{(i)}, c_{2,v}^{(i)}\} \leq r) e_{uv}^{(i)},
\end{equation}
where $d\{c_{1,u}^{(i)},c_{2,v}^{(i)}\}$ is the distance between the $u$-th Type 1 point and the $v$-th Type 2 point in block $i$.

Under the blockwise label-permutation null, the labels of the $n_i$ points in block $i$ are permuted while preserving $m_{i1}$ Type 1 points, $m_{i2}$ Type 2 points, and $n_i - (m_{i1} + m_{i2})$ background points. The analytical expectation and variance of $\hat{K}_{12,i}(r)$ under this null distribution are given in Lemma \ref{lem:biv_moments}.

\begin{lemma} \label{lem:biv_moments}
Under the permutation null distribution, the expectation and variance of $\hat{K}_{12,i}(r)$ for $i=1, \dots, b$ are given by
\begin{align*}
E(\hat{K}_{12,i}(r)) 
&= \frac{|A_i|}{n_i(n_i-1)} R_0^{(i)}, \\
Var(\hat{K}_{12,i}(r)) 
&= \frac{|A_i|^2}{m_{i1}^2 m_{i2}^2} \left\{ R_1^{(i)} h_{1,i} + R_2^{(i)} h_{2,i} + R_3^{(i)} h_{3,i} \right\} - E^2(\hat{K}_{12,i}(r)),
\end{align*}
where
\begin{align*}
h_{1,i} = \frac{m_{i1}m_{i2}}{n_i(n_i-1)}, \quad
h_{2,i} = \frac{m_{i1}m_{i2}(m_{i1}+m_{i2}-2)}{n_i(n_i-1)(n_i-2)}, \quad
h_{3,i} = \frac{m_{i1}m_{i2}(m_{i1}-1)(m_{i2}-1)}{n_i(n_i-1)(n_i-2)(n_i-3)},
\end{align*}
and the quantities $R_0^{(i)}, R_1^{(i)}, R_2^{(i)}$, and $R_3^{(i)}$ are defined as in Lemma \ref{lem:moments}.
\end{lemma}

Using these expressions, we define the blockwise standardized bivariate statistic $Z_{12,i}(r)$ analogously to equation \eqref{eq:standardized_ripley_k}. To aggregate colocalization signals across blocks, we define
\begin{equation}
Z_{12}(r) = \sum_{i=1}^{b} w_{12,i} Z_{12,i}(r),
\end{equation}
where $w_{12,i} = (n_i / p_{12,i})/\sqrt{\sum_{j=1}^b (n_j / p_{12,j})^2}$ with $p_{12,i} = (m_{i1} + m_{i2})/n_i$. Here, $p_{12,i}$ denotes the combined proportion of Type 1 and Type 2 points in block $i$.

For the bivariate analysis, the minimum block size constraints are modified to ensure that each block contains a sufficient number of both point types and background points. Specifically, each block is required to contain at least $\sqrt{m_1}$ Type 1 points, $\sqrt{m_2}$ Type 2 points, and $\min(\sqrt{m_1}, \sqrt{m_2}, \sqrt{n-(m_1+m_2)})$ background points, where $m_1$ and $m_2$ denote the total numbers of Type 1 and Type 2 points in the full observation window, respectively. The asymptotic normal approximation for $Z_{12}(r)$ follows by the same Lyapunov CLT argument in Section~\ref{sec:statistic} and is therefore omitted here.

\section{Simulation studies}
\label{sec:simulation}

This section presents simulation studies evaluating the performance of the proposed spatial clustering test across a range of settings. We compare our method, $B$-$KAMP$, with the original $KAMP$ and $KAMP$ combined with independent thinning at retention probability $\tilde{p}$, which we denote by $KAMP_{\tilde{p}}$ throughout the simulations and real data analysis. Since the simulated point patterns contain tens of thousands of points, we consider $\tilde{p} \in \{0.25, 0.5\}$ for $KAMP_{\tilde{p}}$ to mitigate the computational and memory burden of the original $KAMP$ procedure. 

As an additional scalable baseline, we include a graph-based neighborhood enrichment test developed for high-throughput spatial data, which constructs a radius-based neighborhood graph and assesses spatial clustering through label permutation \citep{palla2022squidpy}. We compute its $p$-values using 1,000 permutations and denote the method by $NET_r$, where $r$ is the radius used to construct the neighborhood graph.

We first describe the data-generating mechanisms under the null and alternative scenarios and then report simulation results in terms of statistical power and computation time.

\subsection{Simulation design}
\label{sec:sim_design}
Let $\lambda_n$ denote the expected total number of points in the observed window, and let $p$ denote the expected abundance of points of interest (POIs). We write $\lambda_m = p\lambda_n$ for the expected number of POIs and $\lambda_n - \lambda_m$ for the expected number of background points. We evaluate the type I error and statistical power under the following null and alternative scenarios. 

Under the null scenario, POIs and background points are generated from two independent homogeneous Poisson processes on $A$ with $\lambda_m$ and $\lambda_n - \lambda_m$, respectively. The two processes are then superimposed to form a single point pattern. This construction represents a setting in which POI labels exhibit no relative clustering with respect to the background point distribution.

Under the alternative scenario, we generate clustered point patterns 
through the following procedure.
\begin{itemize}
    \item \textbf{Step 1: Base pattern generation.} On a bounded 
    window $A$, generate a point pattern from a homogeneous Poisson 
    point process with expected number of points $\lambda_n$.
    
    \item \textbf{Step 2: Identification of cluster centers.} 
    Independently sample $k$ latent cluster centers from the uniform 
    distribution on $A$. These points serve as the centers of 
    potential clustering regions and are not included in the observed 
    point pattern.
    
    \item \textbf{Step 3: Kernel score calculation.} For each point 
    generated in Step~1, compute Gaussian kernel scores with respect 
    to all $k$ latent cluster centers using a prespecified bandwidth 
    parameter $sd$. The score assigned to each point is the maximum of these kernel values.
    
    \item \textbf{Step 4: Probabilistic labeling.} Given its 
    assigned kernel score, each point is independently labeled as a candidate POI through a Bernoulli trial 
    with success probability equal to the kernel score. Points not labeled as candidate POIs are assigned to the background class.
    
    \item \textbf{Step 5: Abundance adjustment.} If the resulting 
    proportion of candidate POIs exceeds the target abundance $p$, excess candidate POIs are randomly relabeled as background points until the empirical POI proportion matches $p$.
\end{itemize}

In our numerical studies, we consider $\lambda_n \in$ $\{20{,}000, 40{,}000, 60{,}000,$ $80{,}000, 100{,}000\}$ and $p \in \{0.1, 0.15, 0.2\}$. Under the null scenario, point patterns are generated on the unit square $[0,1]^2$. Under the alternative scenario, clustered patterns are generated on the larger window $[0,10]^2$ with $k=100$ latent cluster centers and bandwidth $sd = 7$. The larger window is used to accommodate multiple latent cluster centers while avoiding excessive overlap among clustering regions. After generating a point pattern, we construct an adaptive regular grid for block identification by increasing the grid resolution until each grid cell contains at most $\sqrt{n}$ points, resulting in equally sized rectangular grid cells. Throughout the simulations, we set $\rho_1$ equal to the aspect ratio of the observation window and set $\rho_2 = \infty$.

We estimate type I error and power from 500 simulated point patterns under each scenario, using the significance level $\alpha=0.05$. Computational performance is evaluated in terms of runtime using \texttt{R} on a 3.7 GHz AMD Ryzen 5 7500F 6-Core processor. 

To select the evaluation radii, we follow Ripley’s rule of thumb implemented in the \texttt{spatstat} package \citep{baddeley2016spatial}, which recommends using radii between 0 and 0.25 times the shorter side of the observation window over which the $K$-function is computed. Since $B$-$KAMP$ produces blocks with varying sizes and shapes, we define the radius proportionally rather than as a fixed absolute distance. Specifically, simulation results are reported for radii $r \in \{0.05, 0.10, 0.15, 0.20, 0.25\}$ times the corresponding reference length. For $B$-$KAMP$, the reference length is the shorter side of each block, whereas for $KAMP$-based full-window methods and $NET$, it is the shorter side of the full observation window. Therefore, for a given $r$, $B$-$KAMP$ evaluates clustering at a smaller absolute spatial scale than the full-window methods.

\subsection{Simulation results}

Table~\ref{tab:comp_time} reports the average runtimes in seconds for the competing methods across different values of $\lambda_n$. Since the runtimes of $B$-$KAMP$, $KAMP$, and $KAMP_{\tilde p}$ are largely insensitive to the choice of $r$, we report their runtimes for a single representative radius. In contrast, the runtimes of $NET_r$ depend strongly on $r$, and therefore results are reported separately for each radius.
\begin{table}[h!]
\centering
\caption{Comparison of average computation times in seconds across varying $\lambda_n$. Entries marked with ``--'' indicate that the method did not complete due to excessive memory or computational cost.}
\label{tab:comp_time}
\resizebox{\columnwidth}{!}{%
\begin{tabular}{@{}cccccccccc@{}}
\toprule
$\lambda_n$ & $B$-$KAMP$ & $KAMP$ & $KAMP_{0.5}$ & $KAMP_{0.25}$ & $NET_{0.05}$ & $NET_{0.1}$ & $NET_{0.15}$ & $NET_{0.2}$ & $NET_{0.25}$ \\ \midrule
20,000  & 1.5  & 22.6 & 5.0  & 1.3  & 8.1   & 19.5  & 36.4   & 62.7   & 95.7   \\
40,000  & 3.6  & --   & 22.1 & 5.1  & 19.8  & 63.1  & 168.9  & 281.2  & 389.5  \\
60,000  & 7.3  & --   & --   & 13.5 & 39.5  & 136.5 & 435.4  & 646.3  & 897.5  \\
80,000  & 10.6 & --   & --   & 22.5 & 68.5  & 414.0 & 801.9  & 1206.5 & 1803.9 \\
100,000 & 14.8 & --   & --   & 50.2 & 105.8 & 766.2 & 1483.0 & 2170.1 & 3523.8 \\
\bottomrule
\end{tabular}
}
\end{table}

The original $KAMP$ becomes infeasible when $\lambda_n \geq 40{,}000$ because it requires pairwise distance calculations over the full observation window. $KAMP_{0.5}$ encounters the same limitation when $\lambda_n \geq 60{,}000$. Although $KAMP_{0.25}$ remains feasible over the range of $\lambda_n$ considered here, its computational cost still increases rapidly with $\lambda_n$, and the underlying quadratic dependence on the number of retained points is not fundamentally removed by thinning.

$NET_r$ mitigates the full pairwise-distance bottleneck by restricting attention to point pairs that fall within the radius-based neighborhood graph, and it remains executable for all values of $\lambda_n$ considered in this simulation study. However, its computational cost is highly sensitive to both $\lambda_n$ and $r$. As $r$ increases, the neighborhood graph becomes denser, and the permutation-based calibration becomes increasingly expensive. Thus, its rapidly increasing runtime suggests limited scalability for larger or denser spatial point patterns. This limitation is also observed in the real data analysis in Section \ref{sec:real}, where full-window methods, including $NET$, exceed the available computational resources.

In contrast, $B$-$KAMP$ scales favorably across all settings. It remains feasible for all values of $\lambda_n$ and is the fastest method when $\lambda_n \geq 40{,}000$, even compared with $KAMP_{0.25}$, which uses only 25\% of the original points. These results demonstrate the computational scalability of $B$-$KAMP$ for large spatial point patterns.

\begin{figure}[h!]
    \centering
    \includegraphics[width=0.9\textwidth]{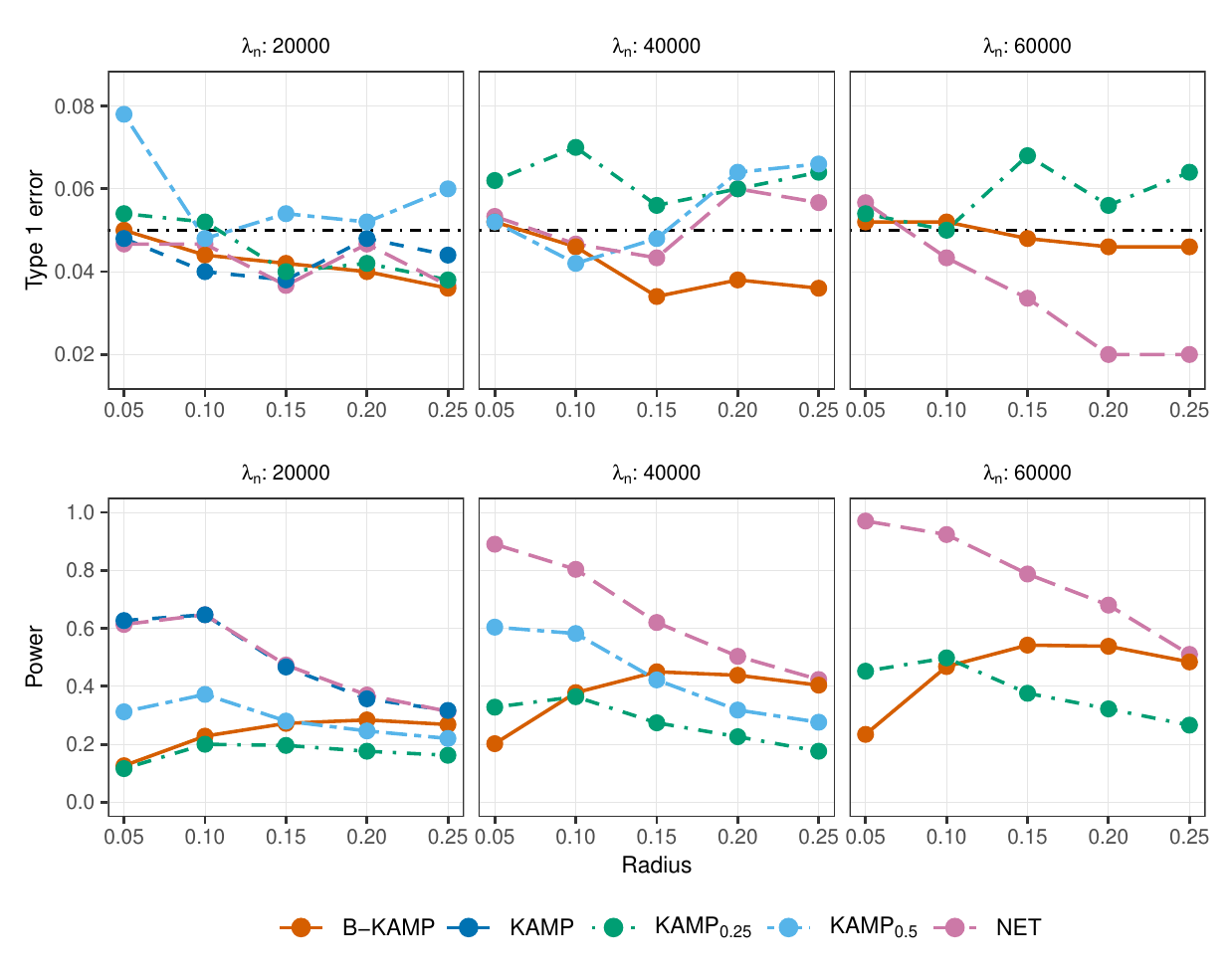}
    \caption{Empirical type I error rate and power across different values of radii $r$ and $\lambda_n$, with a fixed abundance $p=0.1$. The dotted horizontal lines in the top row indicate the nominal level 0.05.}
    \label{fig:simulation}
\end{figure}

Figure \ref{fig:simulation} shows the empirical type I error rate and statistical power across different radii for $\lambda_n \in \{20{,}000,40{,}000,60{,}000\}$ with abundance fixed at $p=0.1$. $KAMP$ and $KAMP_{0.5}$ are omitted for $\lambda_n \geq 40{,}000$ and $\lambda_n = 60{,}000$, respectively, due to their computational limitations, as shown in Table \ref{tab:comp_time}.

The type I error results indicate that $B$-$KAMP$, $KAMP$, and $NET$ control the nominal level reasonably well across all settings. By contrast, $KAMP_{0.25}$ and $KAMP_{0.5}$ show some inflation in type I error, likely due to the additional variability introduced by thinning in these finite-sample settings.

The power results show that $KAMP$ and $NET$ generally achieve the highest power when they are computationally feasible, which is expected because both methods use spatial information from the full observation window. $B$-$KAMP$ therefore should not be viewed as uniformly more powerful than full-window procedures. Rather, its advantage lies in providing a substantially more scalable alternative that retains competitive power at appropriate radii, particularly in settings where full-window $KAMP$ or permutation-based neighborhood methods become computationally impractical.

$B$-$KAMP$ has relatively low power at small radii, especially at $r=0.05$, since such spatial scales are too small to capture clustering signal effectively within the proposed block-based framework (see Section~\ref{sec:sim_design}). As the radius increases, however, the power of $B$-$KAMP$ improves substantially and reaches its highest level around $r=0.15$. At this radius, $B$-$KAMP$ achieves power comparable to $KAMP_{0.5}$ and higher than $KAMP_{0.25}$ across all values of $\lambda_n$. Taken together with the runtime results in Table \ref{tab:comp_time}, these findings indicate that $B$-$KAMP$ provides a favorable balance between statistical power and computational efficiency.

We next examine the effect of POI abundance on computational and statistical performance. Table \ref{tab:comp_time_abundance} reports the average runtimes across abundance levels $p \in \{0.1,0.15,0.2\}$ with $\lambda_n=40{,}000$. Across all abundance levels, $B$-$KAMP$ is the fastest method. Furthermore, its runtime decreases as abundance increases, whereas no clear monotone pattern is observed for the competing methods. This trend is consistent with the behavior of the adaptive spatial blocking algorithm: when POIs are more abundant, valid blocks satisfying the point-count constraints can be identified more efficiently, resulting in smaller blockwise computations.

\begin{table}[h!]
\centering
\caption{Comparison of average computation times in seconds across varying abundance levels with $\lambda_n=40,000$.}
\label{tab:comp_time_abundance}
\resizebox{\columnwidth}{!}{%
\begin{tabular}{@{}ccccccccc@{}}
\toprule
Abundance & $B$-$KAMP$ & $KAMP_{0.5}$ & $KAMP_{0.25}$ & $NET_{0.05}$ & $NET_{0.1}$ & $NET_{0.15}$ & $NET_{0.2}$ & $NET_{0.25}$ \\ \midrule
0.10 & 3.7 & 22.1 & 5.1 & 19.8 & 63.3 & 170.0 & 286.0 & 391.0 \\
0.15 & 3.1 & 21.7 & 5.2 & 20.1 & 62.1 & 176.9 & 283.8 & 392.0 \\
0.20 & 2.7 & 21.7 & 4.9 & 19.5 & 59.2 & 164.0 & 271.0 & 371.0 \\
\bottomrule
\end{tabular}
}
\end{table}

\begin{figure}[h!]
\centering
\includegraphics[width=0.9\textwidth]{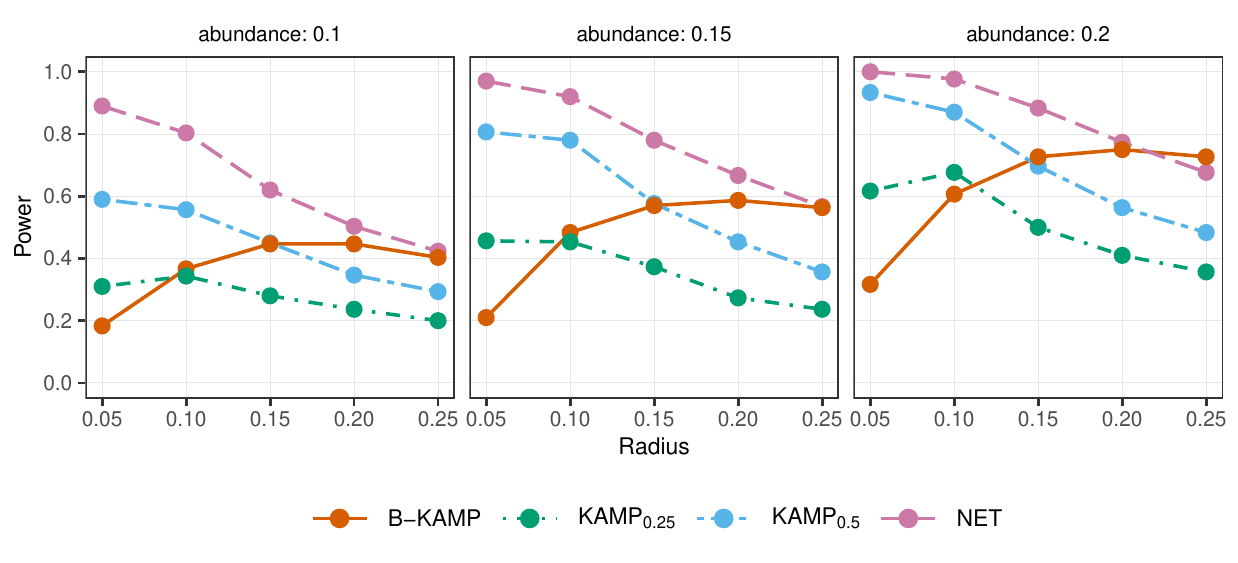}
\vspace{-0.5em}
\caption{Power comparison across different radii and abundance levels at $\lambda_n = 40,000$.}
\label{fig:simulation_abundance}
\end{figure}

Figure \ref{fig:simulation_abundance} reports the corresponding power curves. Consistent with the results in Figure~\ref{fig:simulation}, at $r = 0.15$, $B$-$KAMP$ demonstrates comparable power to $KAMP_{0.5}$ and outperforms $KAMP_{0.25}$ across all abundance levels.

Overall, $B$-$KAMP$ offers a scalable blockwise alternative that maintains valid type I error control, achieves competitive power at suitable radii, and remains computationally practical in regimes where full-window $KAMP$, heavily thinned $KAMP$, or permutation-based neighborhood methods face clear limitations.

\section{Applications to high-throughput spatial proteomics}
\label{sec:real}

\subsection{Data}

We analyze a publicly available spatial proteomics dataset of healthy human intestine tissue generated using co-detection by indexing (CODEX) multiplexed imaging \citep{hickey2022processed, hickey2023organization}. This dataset was originally collected to characterize the cellular composition and spatial organization of the human intestine and to provide a healthy reference for future single-cell spatial studies. It consists of 64 tissue sections collected from 8 donors across 8 distinct anatomical regions, including the small intestine (duodenum, proximal jejunum, mid-jejunum, and ileum) and the large intestine (ascending, transverse, descending, and sigmoid colon). In total, the dataset contains 66 images, including two replicate tissue sections. After image processing and cell segmentation, the data are provided in tabular form and contain approximately 2.6 million individual cells, with each image containing tens of thousands of cells. For each cell, the dataset records two-dimensional spatial coordinates within the tissue section together with the annotated cell type. Further details on tissue acquisition and image preprocessing are provided in \citet{hickey2022processed} and \citet{hickey2023organization}.

Motivated by multicellular neighborhood analysis of \citet{hickey2023organization}, we focus on two spatial patterns: the univariate clustering of plasma cells and the bivariate colocalization between plasma cells and macrophages. \citet{hickey2023organization} reported a Plasma-Cell-Enriched multicellular neighborhood characterized by high plasma cell density and co-enrichment with antigen-presenting cells, including macrophages, across intestinal regions. These findings were obtained through neighborhood clustering analyses, which provide exploratory summaries of cellular spatial proximity but are not designed as formal statistical tests of clustering or colocalization. We therefore apply the proposed block-based framework to test two spatial patterns across the images in the dataset. One image containing no plasma cells is excluded, leaving 65 images for subsequent analysis.
 
\begin{figure}[h!]
    \centering
    \includegraphics[width=0.7\textwidth]{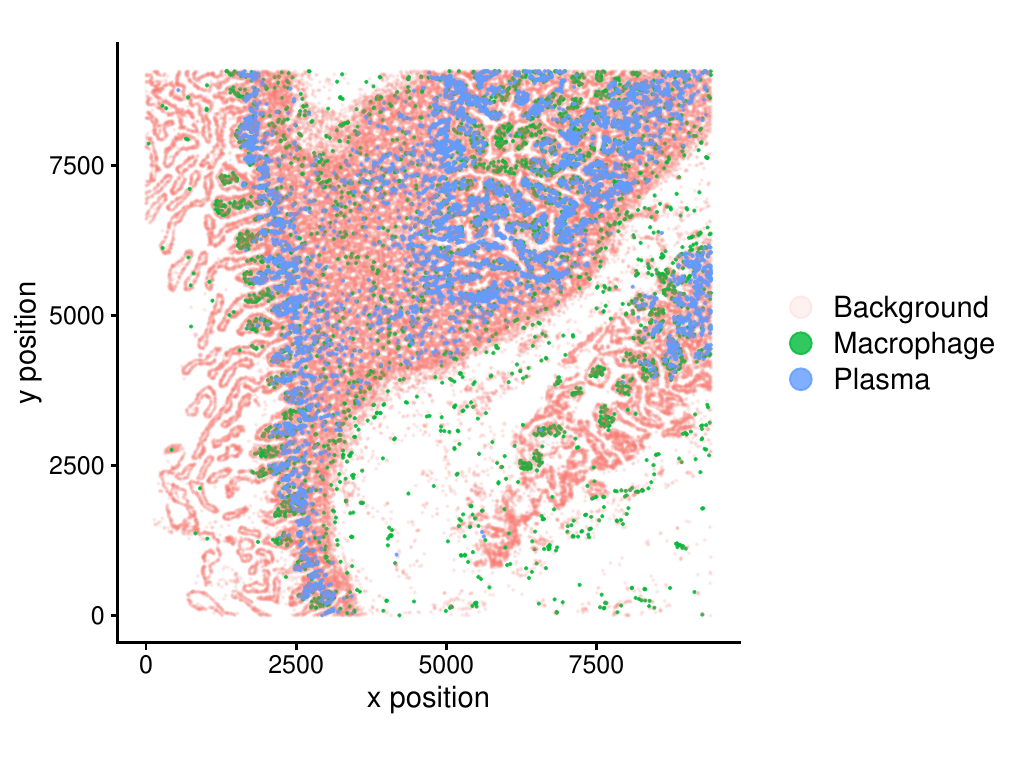}
    \vspace{-1em}
    \caption{Illustration of the cell organization in the selected tissue image.}
    \label{fig:cell_dist}
\end{figure}

For illustration, Figure~\ref{fig:cell_dist} shows the spatial distribution of plasma cells and macrophages in the tissue section containing the largest number of cells. This image consists of 7,192 plasma cells, 3,806 macrophages, and 78,909 background cells, comprising all other cell types, for a total of 89,907 observed cells within the observation window $[2, 9406] \times [2, 9070]$.

\subsection{Parameter selection}

\begin{figure}[h!]
    \centering
    \includegraphics[width=0.85\textwidth]{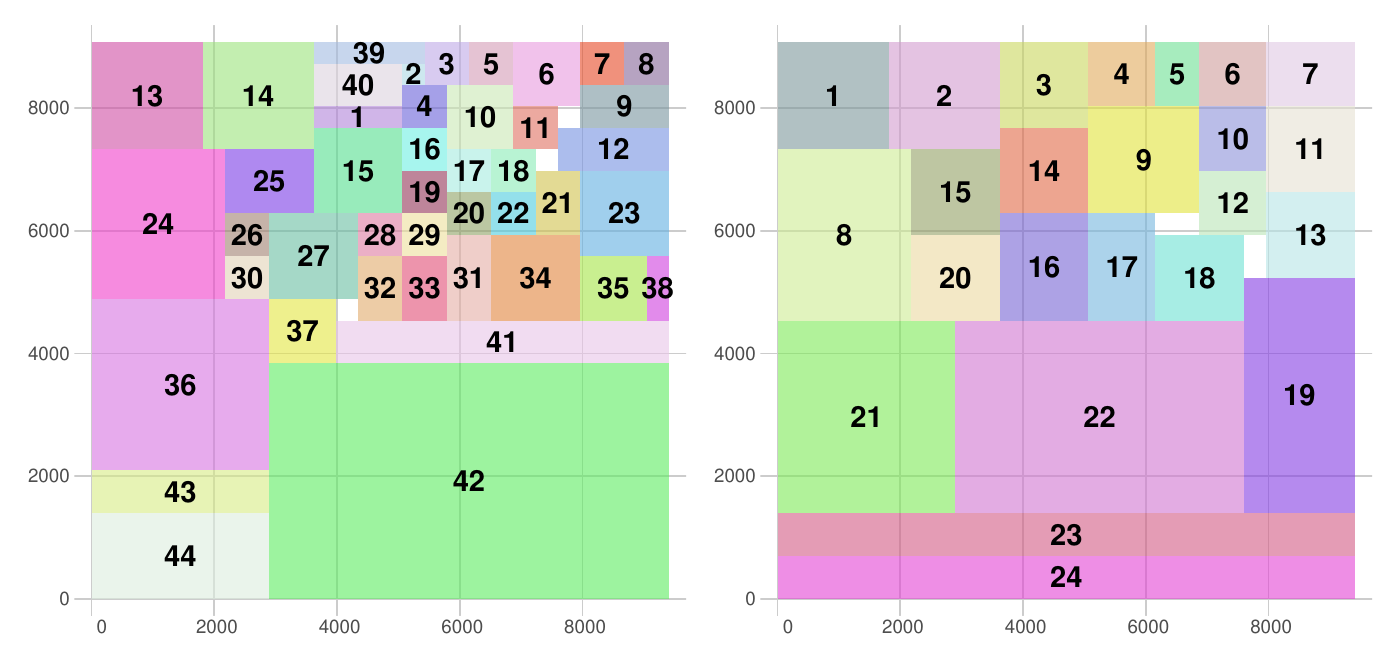}
    \caption{Resulting blocks from the proposed blocking algorithm applied to the image in Figure~\ref{fig:cell_dist}. The left panel displays the blocks for the univariate analysis of plasma cells, and the right panel shows the bivariate analysis of plasma cells and macrophages.}
    \label{fig:real_blocks}
\end{figure}

For each image, we first construct an adaptive regular grid over the observed window described in Section~\ref{sec:sim_design}. Specifically, the grid resolution is increased until each grid cell contains at most $\sqrt{n}$ points, where $n$ denotes the total number of cells in the image. We then apply the adaptive spatial blocking algorithm described in Section~\ref{sec:total_procedure}. Throughout the real data analysis, we set the Phase I parameter $\rho_1$ equal to the aspect ratio of the original image and the Phase II parameter $\rho_2$ to $\infty$. Figure~\ref{fig:real_blocks} shows resulting block configurations for the image displayed in Figure~\ref{fig:cell_dist}. The left panel corresponds to the univariate analysis of plasma cells, whereas the right panel corresponds to the bivariate analysis of plasma cells and macrophages.

Due to the computational limitations of the full-window methods, we consider $KAMP_{\tilde{p}}$ with $\tilde{p} \in \{0.1, 0.25\}$ as competing methods. The full-window methods, including $KAMP$, $KAMP_{0.5}$, and $NET$, are excluded because their computations exceed the available memory limit in this dataset. The evaluation radii for the $K$-function are set to 5 equally spaced values from 0.05 to 0.25, defined relative to the shorter side of the window over which the $K$-function is computed.

\subsection{Results}

\begin{figure}[t!]
    \centering
    \includegraphics[width=0.99\textwidth]{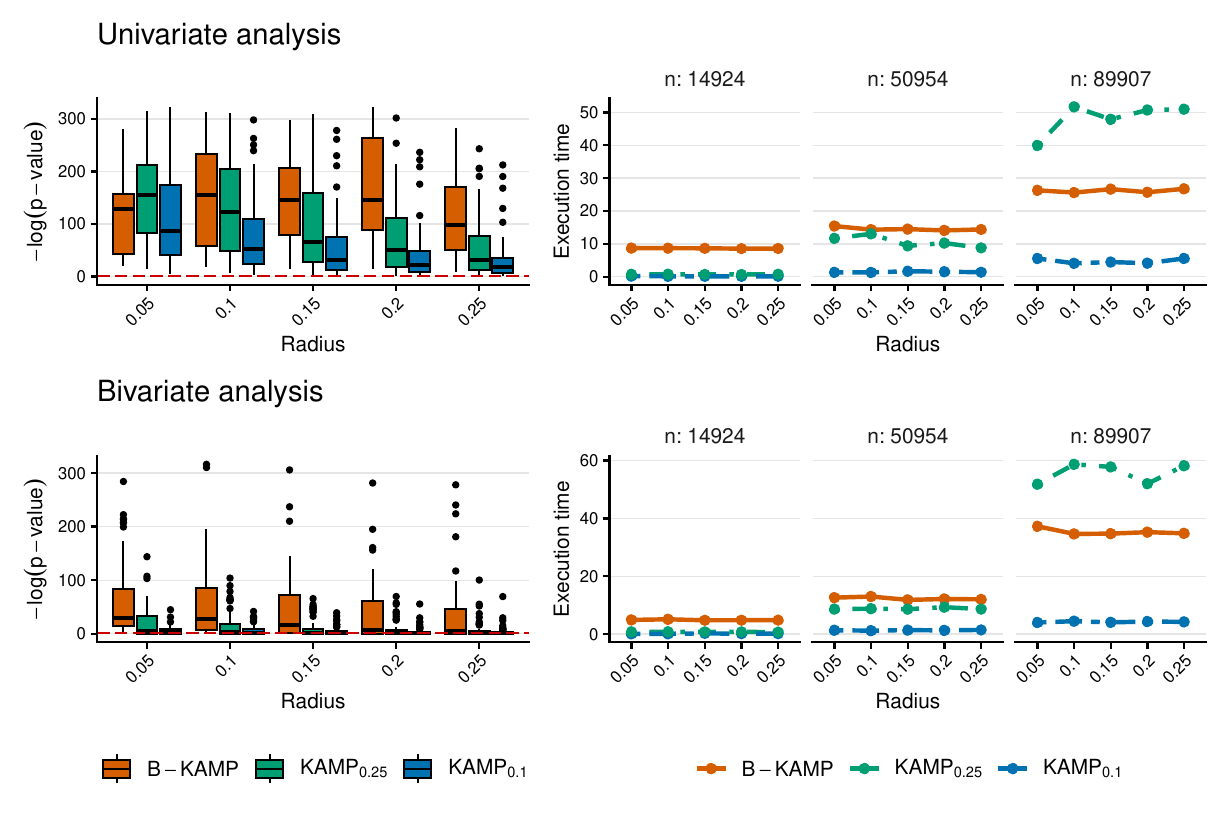}
    \vspace{-0.5em}
    \caption{Performance comparison of $B$-$KAMP$ and baseline methods for univariate plasma cell clustering and bivariate plasma cell-macrophage colocalization. The left panels show $-\log(p\text{-value})$ across 65 images over different radii with the red horizontal line indicating $-\log(0.05)$. The right panels show execution times in seconds for representative small, moderate, and large images.}
    \label{fig:plot_bivariate}
\end{figure}

Figure~\ref{fig:plot_bivariate} compares $B$-$KAMP$ with computationally feasible thinned $KAMP$ in terms of statistical evidence and computation time. The left panels show boxplots of $-\log(p\text{-value})$ across the 65 images, with the red horizontal line indicating $-\log(0.05)$. Larger values therefore correspond to stronger evidence against the null hypothesis. The right panels report execution times in seconds for three representative images with $n \in \{14{,}924,\ 50{,}954,\ 89{,}907\}$ total cells, referred to as small, moderate, and large images, respectively. All runtimes are evaluated in \texttt{R} on a 3.7 GHz AMD Ryzen 5 7500F 6-Core processor. The top row corresponds to the univariate clustering analysis of plasma cells, and the bottom row corresponds to the bivariate colocalization analysis of plasma cells and macrophages.

In the univariate analysis, $KAMP_{0.1}$ has the shortest runtime due to its aggressive subsampling, but it tends to yield larger $p$-values than other methods. $B$-$KAMP$ tends to produce smaller $p$-values than $KAMP_{0.25}$ across most radii, except $r = 0.05$. In particular, $B$-$KAMP$ rejects the null hypothesis of no plasma cell clustering for all images across all radii, whereas $KAMP_{0.25}$ fails to reject for some images at $r = 0.2$ and $r = 0.25$. In terms of computation time, $KAMP_{0.25}$ is faster than $B$-$KAMP$ for the small and moderate images. For the large image, however, $B$-$KAMP$ scales more favorably and becomes substantially faster than $KAMP_{0.25}$.

The bivariate analysis shows a similar pattern. $B$-$KAMP$ generally provides stronger evidence of colocalization than the baseline methods across all radii, whereas $KAMP_{0.1}$ again yields comparatively larger $p$-values due to more aggressive subsampling. In terms of runtime, $B$-$KAMP$ outperforms $KAMP_{0.25}$ for the large image.

Overall, the real data analysis suggests strong spatial aggregation of plasma cells and colocalization between plasma cells and macrophages in the healthy human intestine dataset. These findings provide formal statistical support for the spatial patterns reported in the neighborhood-based analysis of \citet{hickey2023organization}, while demonstrating the scalability of the proposed block-based framework for large spatial proteomics images.

\section{Discussion}
\label{sec:discussion}

In this paper, we proposed the adaptive spatial blocking algorithm for scalable spatial clustering inference in high-throughput spatial data. By incorporating explicit block constraints, multi-phase identification, and a complement step for residual grid cells, the algorithm constructs a collection of disjoint rectangular blocks while reducing the loss of spatial coverage, thereby enabling fast and reliable spatial clustering inference. Building on this algorithm, we develop $B$-$KAMP$ by combining the proposed blocking strategy with the $KAMP$ framework. Simulation studies demonstrated that $B$-$KAMP$ provides a favorable balance between statistical power and computational efficiency. In the analysis of healthy human intestine spatial proteomics data, the proposed framework provides consistent evidence of spatial aggregation of plasma cells and colocalization between plasma cells and macrophages, while scaling favorably to large images.

Several directions for further development remain. First, the computational efficiency and statistical behavior of the proposed algorithm depend on the initial grid resolution. An overly fine grid increases the cost of the block identification and complement steps, whereas an overly coarse grid may yield too few blocks, each containing a large number of points. Developing a data-driven strategy for selecting the grid resolution could improve the stability and usability of the algorithm. Second, the current block complement step (Phase III) adopts a greedy one-step procedure that expands residual grid cells into adjacent blocks based on local geometric criteria. Future work could consider global optimization or multi-step lookahead strategies to better control the overall aspect ratio of the resulting block collection and prevent certain blocks from becoming disproportionately large or elongated.

The proposed framework is interpreted as a scalable conditional blockwise test rather than a full-window permutation distribution of the original $KAMP$ statistic. By performing inference within disjoint blocks, the method reduces computational cost but does not use spatial relationships across block boundaries. As a result, clustering patterns that extend across block boundaries may be attenuated, and some loss of power relative to full-window methods is expected. Developing boundary-aware aggregation rules, dependence-adjusted combinations of block-level statistics, and more refined weighting schemes for different clustering regimes would be valuable extensions.

Finally, as spatial multi-omics technologies continue to advance, spatial data may increasingly include three-dimensional tissue imaging and 3D point patterns. Extending the proposed two-dimensional blocking algorithm to three dimensions, for example through rectangular voxels, would be a natural next step toward scalable spatial inference for more complex spatial data.

\clearpage
\bibliography{references}

\end{document}